\documentclass[aoas,MSNbibl,nameyear,dvips]{arximspdf}
\usepackage{graphicx}

\doi{10.1214/12-AOAS563} 
\volume{6}
\issue{4}
\pubyear{2012}
\firstpage{1707}
\lastpage{1729}

\makeatletter
\newcommand{\mathbfo}{}

\newcommand{\bmm}{\bolds}
\newcommand{\bmb}{\mathbf}
\newcommand{\mbfs}{\bolds}
\newcommand{\bbeta} {\bolds{\beta}}
\newcommand{\bpsi} {\bolds{\psi}}
\newcommand{\bphi} {\bolds{\phi}}
\newcommand{\bgamma} {\bolds{\gamma}}
\newcommand{\bchi} {\bolds{\chi}}
\newcommand{\bdelta}{\bolds{\delta}}
\newcommand{\etab} {\bolds{\eta}}
\newcommand{\mS} {\mathcal{B}}

\makeatother

\begin{document}
\begin{frontmatter}

\title{Toxicity profiling of engineered nanomaterials via multivariate
dose-response surface modeling}
\runtitle{Toxicity profiling of engineered nanomaterials}

\begin{aug}
\author[a]{\fnms{Trina} \snm{Patel}\corref{}\ead[label=e1]{trpatel@ucla.edu}},
\author[a]{\fnms{Donatello} \snm{Telesca}},
\author[b]{\fnms{Saji}~\snm{George}}
\and
\author[b]{\fnms{Andr\'{e}~E.}~\snm{Nel}}
\runauthor{Patel, Telesca, George and Nel}
\affiliation{University of California, Los Angeles}
\address[a]{T. Patel\\
D. Telesca\\
Department of Biostatistics\\
School of Public Health\\
University of California, Los Angeles\\
Los Angeles, California 90095-1772\\
USA\\
\printead{e1}} 
\address[b]{S. George\\
A. E. Nel\\
Department of Medicine\\
Division of NanoMedicine\\
University of California, Los Angeles\\
Los Angeles, California 90095-1772\\
USA\\
and\\
California NanoSystems Institute\\
University of California, Los Angeles\\
Los Angeles, California 90095-1772\\
USA}
\end{aug}

\received{\smonth{3} \syear{2012}}

%
\begin{abstract}
New generation {in vitro} high-throughput screening (HTS) assays
for the assessment of engineered nanomaterials provide an opportunity
to learn how these particles interact at the cellular level,
particularly in relation to injury pathways. These types of assays are
often characterized by small sample sizes, high measurement error and
high dimensionality, as multiple cytotoxicity outcomes are measured
across an array of doses and durations of exposure. In this paper we
propose a probability model for the toxicity profiling of engineered
nanomaterials. A hierarchical structure is used to account for the
multivariate nature of the data by modeling dependence between outcomes
and thereby combining information across cytotoxicity pathways. In this
framework we are able to provide a flexible surface-response model that
provides inference and generalizations of various classical risk
assessment parameters. We discuss applications of this model to data on
eight nanoparticles evaluated in relation to four cytotoxicity parameters.
\end{abstract}

%

\end{frontmatter}

\section{Introduction}\label{sec1}
Nanotechnology is rapidly growing and currently used in various
industries such as food, agriculture, electronics, textiles and health
care. The widespread use of engineered nanomaterials (ENM) in over 800
consumer products increases the likelihood that these materials will
come into contact with humans and the environment [\citet{Maynard2006},
\citet{Kahru2009}]. Many biological processes take place at the
nanoscale level,
and the introduction of ENMs into living organisms could lead to
interference in the molecular and cellular processes that are critical
to life [\citet{Nel2009}]. This potential for human and environmental
hazard has spurred recent interest in early identification of
potentially hazardous nanomaterials. Knowledge about the potential
hazard of nanomaterials is still lacking and a lot of study is required
to understand how ENM properties
such as size, shape, agglomeration state, solubility and surface
properties could lead to hazard generation at the nano-bio interface
[\citet{Stern2008}, \citet{Nel2006}].

\begin{figure}

\includegraphics{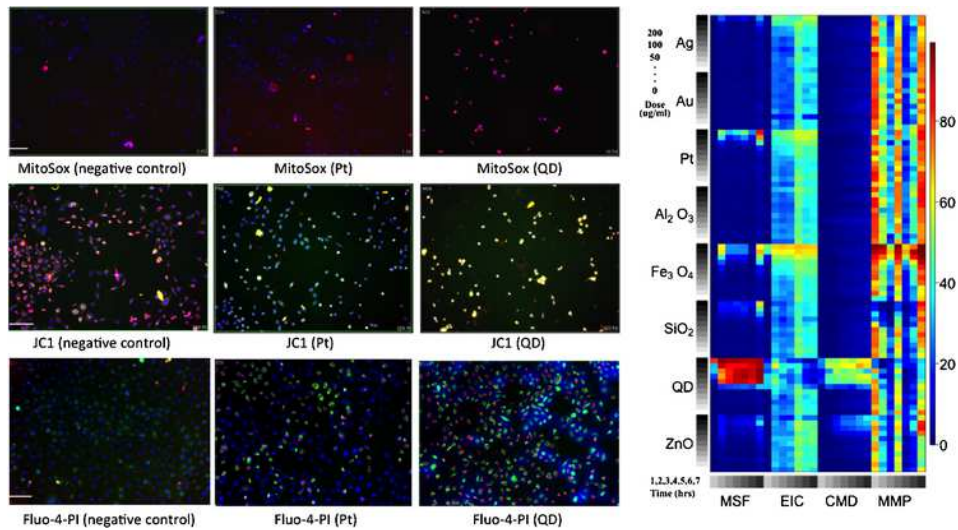}

\caption{Fluorescence images and heat map of raw data.
On the left are fluorescence images of RAW cells treated with various
nanomaterials (quantum dot, platinum and a negative control) and dyed
with compatible dye combinations including MitoSox, JC1, PI and Fluo-4.
The subsequence fluorescence read-out, measured at varying wavelengths,
provides a measure of the number of cells positive for the response.
On the left is a heat map of the raw data for each particle and outcome.
Colder colors indicate a smaller percentage of cells positive for the
response and warmer colors indicate a larger percentage of cells
positive for the response.}
\label{figraw}
\end{figure}

Current research in nano-toxicology includes new generation
high-through\-put screening (HTS) assays, which enable the simultaneous
observation of multiple cellular injury pathways across an array of
doses and times of exposure. In this article, for example, we analyze
data on eight metal and metal oxide nanoparticles, monitored in
relation to four cellular injury responses, derived from the
hierarchical oxidative stress model of \citet{Nel2006} and \citet
{Xia2006}. All four outcomes are measured contemporaneously over a grid
of ten doses and seven hours of exposure (see Figure~\ref{figraw}).
The four measured responses include mitochondrial superoxide formation,
loss of mitochondrial membrane potential, elevated intracellular
calcium and membrane damage [\citet{GeorgePokhrel2010}]. For increasing
dosage and duration of exposure, we observe typical dose-response kinetics,
with outcomes possibly depending on one another.

These assays provide an opportunity to help define biological
relationships and may suggest which nanoparticles are likely to have an
in vivo effect. While HTS assays cannot replace traditional animal
studies, they are less costly, less labor intensive and can be used to
explore the large number of potential nanomaterial variables that can
influence human health hazards [\citet{Meng2010}, \citet{Shaw2008},
\citet
{Maynard2006}]. The feasibility and utility of HTS assays have been
illustrated in various fields such as functional genomics, with the use
of microarray technology, as well as in pharmacology for the rapid
screening of potential drug targets [\citet{Hoheisel2006}, \citet{White2000}].
In toxicology, risk assessment involves the characterization of hazard
as well as the potential for exposure while accounting for all
assumptions and uncertainties. The HTS framework provides a wealth of
information about cellular injury pathways but proves a challenge for
the classic risk assessment paradigm. In fact, there is still
disagreement in the HTS setting on how to define and how appropriate
are classical risk assessment parameters such as no observable adverse
effect level (NOAEL), the lowest observable adverse effect level
(LOAEL) and the dose that produces 50\% of the maximum response (EC50),
among others.

Parametric functions such as families of sigmoidal curves are
frequently used to fit dose-response data. Some commonly used sigmoidal
models include log-logistic models, log-normal models and Weibull
models [see \citet{Ritz2010} for a recent review of these models]. The
log-logistic functions are the most frequently used for modeling
dose-response data in toxicology. The four parameter log-logistic model
can be expressed as follows:
%
\begin{equation}
\label{eqlog-logistic} %
f (x; b, c, d, h) = c + \frac{d-c}{1+\exp[b\{\log
(x)-\log(h)\}]}.
\end{equation}
Here $h$, the inflection point in the curve, provides a convenient risk
assessment parameter, since it can be interpreted as the 50\% effective
or inhibitory dose (EC50, IC50) [\citet{Emmens1940}]. Other special
cases of this model include the 3 parameter log-logistic model which
leads to the famous Hill equation [\citet{Hill1910}] and special cases
of the Michaelis--Menton kinetics. Further extensions of these models
include the five parameter log-logistic function, which provides a bit
more flexibility by allowing the function to be asymmetric [\citet
{Finney1979}], and the Brain-Cousens model, which includes an extra
parameter to account for a possible favorable response to a toxin at
low concentrations [\citet{Calabrese2003}]. In general, these models
assume that the dose-response function is completely known apart from
the few parameters to be estimated, usually by determining which values
of the parameters result in the best fit to the dose-response function.

Several other methods have been proposed to model nonlinear
dose-response relationships relaxing strictly parametric assumptions.
\citet{Ramasay1988} proposed the use of monotone regression splines to
model a dose-response function. In this case, piecewise polynomials or
splines can allow greater flexibility while achieving monotonicity by
imposing constraints on the estimated function. \citet{li2004} proposed
the use of linear B-splines with one random interior knot to model a
nonlinear dose-response curve. In this context, the random interior
knot provides inference on the dose at which the toxin begins to take
effect and thereby provides a useful parameter for risk assessment.
\citet{Kong2006} suggested the use of functions that combine smoothing
spline techniques and the nonnegativity properties of cubic B-splines
to estimate the dose-response curve. The use of nonparametric
techniques to estimate dose-response curves often provides a more
realistic representation
of the data generating process. At the same time, however, some of
these techniques make it more difficult to interpret the model in terms
of classical risk assessment.

Recent literature advocates the simultaneous use of multiple outcomes
to assess risk. \citet{Regan1999} proposed a bivariate dose-response
model that accounts for the dependence among outcomes of developmental
toxicity using generalized estimating equations. \citet{Geys2001}
proposed a similar model for risk assessment of developmental toxicity,
but approached the problem using latent variables. \citet{Yu2005}
suggested a model for quantitative risk assessment of bivariate
continuous measures of neurotoxicity using percentile regression. These
methods are often aimed at the analysis of one potentially toxic agent
as it relates to adverse events or continuous outcomes observed in
association with exposure over a range of doses. Their direct
applicability to the general HTS setting described earlier is therefore limited.

From a statistical perspective, cellular interrogation data based on
high-throughput platforms
can be characterized as multivariate dependent observations. Each
nanoparticle is indeed
associated with a multiple set of cellular outcomes recorded both
longitudinally, in relation to different exposure
durations, and cross-sectionally, in relation to a dose escalation
design. This particular design structure suggests
that valid statistical inference must account for potentially complex
patterns of dependence between
different observations. A reasonable dependence scheme might, for
example, assume data to be
dependent within outcome and particle, as well as between outcomes for
the same particle.

In conjunction with considerations related to the joint sampling
distribution of these data structures,
appropriate statistical treatment must account for nonlinearities in
the mean response associated
with dose and duration dynamics. While, in principle, one can choose to
define a random response surface
in a completely nonparametric fashion, it is important to maintain a
certain degree of interpretability, especially in relation
to standard hazard assessment quantities of interest to substantive
scientists. In summary, perhaps reductively, the overall modeling
challenge lies in the definition of a flexible and interpretable
probabilistic representation for
a family of dependent dose-response random surfaces.

In this paper we propose a hierarchical dose-response model for the
analysis of HTS data from nanotoxicology.\vadjust{\goodbreak} Our model builds on earlier
work [\citet{Hastie1986}, \citet{li2004}], expanding on them to account
for the multivariate nature of the data and to address the estimation
of a series of two-dimensional dose-response surfaces. We provide a
flexible framework for modeling dose and duration response kinetics
jointly, while providing inference on several risk assessment
parameters of interest. We utilize a hierarchical structure to define
dependence between outcomes and thereby borrow strength across injury
pathways, providing the basis for a comprehensive risk assessment
paradigm in HTS studies. We account for outlying observations via a
$T$-distributed error model and describe how to carry out inference for
the model parameters and
their functions on the basis of simulated draws from their posterior
distribution. To our knowledge, we are
the first to propose a principled statistical methodology for the joint
analysis of this new generation of {in vitro} data.

The remainder of the article is organized as follows. In Section~\ref
{secModel} we introduce the proposed model. In Section~\ref{subsecestinf} we discuss
parameter estimation and associated inferential details. Section~\ref
{secApplication} employs the proposed model for the analysis of~8 metal
oxide nanomaterials and describes inference for various risk assessment
parameters of interest. We conclude with a critical discussion of the
limitations and possible extensions of our method in Section~\ref
{secDiscussion}.

\section{Model formulation}\label{secModel}
\subsection{Model description} \label{subsecDescription}
In this section we describe a dose-response model for a general HTS
study, where we monitor a multivariate continuous outcome~$y$,
corresponding to $J$ cytotoxicity parameters, in association with the
exposure of a number of cells to $I$ different ENMs. More precisely,
let $y_{ijk}(d,t)$ denote a multivariate response corresponding to ENM
$i$ ($i=1,\ldots,I$), cytotoxicity parameter $j$ ($j=1,\ldots,J$) and
replicate $k$ ($k=1,\ldots,K$) at dose $d\in[0,D]$ and time $t\in[0,T]$.
In typical applications one observes
$y$ over a discrete set of doses $\tilde{d}=(d_1, \ldots, d_{m_1})'$ and
exposure times $\tilde{t}=(t_1, \ldots. , t_{m_2})'$. However,
for clarity of exposition, we simplify our notation and without loss of
generality refer to a general dose $d\in[0, D]$ and time $t\in[0,T]$.
We introduce the following 4-stage hierarchical model.

\textit{Stage \textup{1:} Sampling model}.
The observed response of particle $i$, cytotoxicity parameter $j$ and
replicate $k$ is modeled as
%
\begin{equation}
\label{eqstage1} y_{ijk}(d,t) = m_{ij}(d,t) +
\varepsilon_{ijk}(d,t),
\end{equation}
where $ \varepsilon_{ijk}(d,t) \sim N(0,\sigma_{\varepsilon_j}^{2}/\tau_{i}) $.
Here $m_{ij}(d,t)$ denotes the response surface for particle $i$ and
outcome $j$. The proposed response surface describes
dose and duration kinetics for all $d\in[0,D]$ and $t\in[0,T]$ and is
expected to exhibit a nonlinear dynamic over these domains. The
distribution of $y_{ijk}$ is modeled in terms of the error term
$\varepsilon_{ijk}$ as a scaled mixture of normal random variables to
account for outlying observations. The error variance is defined in
terms of the measurement error variance $\sigma^2_{\varepsilon_j}$,
specific to cytotoxicity parameter $j$, and on ENM-specific variance
inflation parameter $\tau_{i}$. If we define the joint distribution of
$\varepsilon_{ijk}(d,t)$ and $\tau_{i}$ as $P(\varepsilon_{ijk}(d,t),\tau_{i})=P(\varepsilon_{ijk}(d,t)\mid\tau_{i},\sigma_{\varepsilon_j})P(\tau_{i}\mid\nu)$, choosing $\varepsilon_{ijk}(d,t)\mid\tau_{i},\sigma_{\varepsilon_j} \sim N(0,\sigma_{\varepsilon_j}^{2}/\tau_{i})$ and $\tau_{i}\mid\nu\sim \operatorname{Gamma}(\nu/2,\nu/2)$, it can be shown that the marginal
density of $\varepsilon_{ijk}(d,t)\mid\sigma_{\varepsilon_j}^{2}$ is
distributed as a $T(\sigma_{\varepsilon_j}^{2},\nu)$ [\citet{West1984}].
Under this framework, we can borrow strength across all ENMs by
assuming the error variance is the same, but retain robustness in the
model by allowing ENM-specific departures from normality. We allow the
measurement error $\sigma_{\varepsilon_j}$ to vary between cytotoxicity
parameters due to heterogeneity in the cytotoxicity outcomes.

\textit{Stage \textup{2:} Response model at the ENM by cytotoxicity
parameter level}.
The dose-response surface $m_{ij}(d, t)$ spans two dimensions (dose and
time), and is modeled in an additive fashion as described by \citet{Hastie1986}.
If we let $(\alpha_{ij}, \bbeta^\prime_{ij}, \bphi^\prime_{ij},
\bgamma_{ij}^\prime,$ $\bpsi_{ij}^\prime, \bdelta_{ij}^\prime, \bchi_{ij}^\prime)^\prime$
be a parameter vector indexing the dose-response surface $m_{ij}(d,t)$,
we can then define
%
\begin{equation}
\label{eqstage2} m_{ij}(d,t) = \alpha_{ij} +
f_{ij}(d; \bphi_{ij},\bbeta_{ij}) +
g_{ij}(t; \bpsi_{ij}, \bgamma_{ij}) +
h_{ij}(d,t; \bchi_{ij}, \bdelta_{ij}).
\end{equation}
Here $f_{ij}(d; \bphi_{ij}, \bbeta_{ij})$ is a function modeling the
effect of dose $d$ on response $j$ for ENM $i$. Similarly, $g_{ij}(t;
\bpsi_{ij}, \bgamma_{ij})$ is the function modeling the effect of time
$t$ and $h_{ij}(d,t; \bchi_{ij}, \delta_{ij})$ is the function
modeling the interactive effect of dose and time. More specifically, we
model the interaction of dose and time in a semi-parametric fashion as
$h_{ij}(dt; \bchi_{ij}, \bdelta_{ij})$. This parameterization allows
us to retain direct interpretation of the model parameters, while
avoiding over-fitting of sparse data.
To ensure likelihood identifiability, we require, without loss of
generality, that $f_{ij}(d=0;  \bphi_{ij}, \bbeta_{ij}) = 0$,
$g_{ij}(t=0; \bpsi_{ij}, \bgamma_{ij}) = 0$, and $h_{ij}(dt=0;
\bchi_{ij}, \bdelta_{ij}) = 0$. The parameters $\alpha_{ij}$ can therefore
be interpreted as the background response level for each particle and outcome.

\begin{figure}

\includegraphics{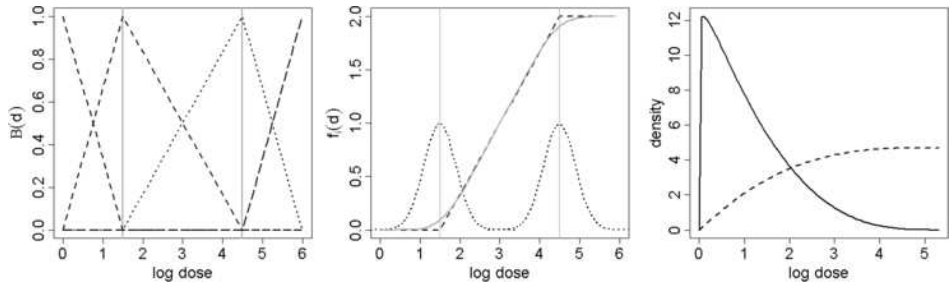}

\caption{Dose-response
as a change-point model. (\textit{left}) B-spline basis function of
degree 1, corresponding to change points (interior knots) at log doses
of 1.5 and 4.5. (\textit{Middle}) Example dose-response curve. The basis
function on the left corresponds to a spline function with 2 change
points. Each random change point has a corresponding distribution,
resulting in a smooth dose-response curve. (\textit{Right}) Example of a
marginal prior distribution on the change points corresponding to the
dose-response curve on the left. This formulation favors (a priori) the
choice of conservative values for the location of the first change
point (\textit{solid line}), and a relatively diffuse prior for our
second change point (\textit{dotted line}).}
\label{figknots}
\end{figure}

We model dose-response curves $f_{ij}(d; \bphi_{ij}, \bbeta_{ij})$,
duration-response curves $g_{ij}(t; \bpsi_{ij}, \bgamma_{ij})$ and
dose-time response curves $h_{ij}(dt; \bchi_{ij}, \bdelta_{ij})$ as
linear combinations of basis functions. Specifically, we use linear
B-splines with two random interior knots as points where the slope
changes in a piecewise linear fashion. Let $\mS(x,\etab)$ denote a
4-dimensional B-spline basis with interior knots $\etab=(\eta_{1},\eta_{2})^\prime$. Also, let $\bbeta_{ij} = (\beta_{ij1},\ldots,\beta_{ij4})^\prime$, $\bgamma_{ij} = (\gamma_{ij1}, \ldots,\gamma_{ij4})^\prime
$ and $\bdelta_{ij} = (\delta_{ij1}, \ldots,\delta_{ij4})^\prime$ be
$4$-dimensional vectors of spline coefficients. The functions
$f_{ij}(d; \bphi_{ij}, \bbeta_{ij})$, $g_{ij}(t; \bpsi_{ij},
\bgamma_{ij})$ and $h_{ij}(dt; \bchi_{ij}, \bdelta_{ij})$ can then be
represented as follows:
%
\begin{eqnarray}
\label{eqstage2ctd} %
f_{ij}(d; \bphi_{ij},
\bbeta_{ij}) &=& \mS(d,\bphi_{ij})^\prime
\bbeta_{ij},
\nonumber
\\
g_{ij}(t; \bpsi_{ij}, \bgamma_{ij}) &=& \mS(t,
\bpsi_{ij})^\prime \bgamma_{ij},
\\
h_{ij}(dt; \bchi_{ij}, \bdelta_{ij}) &=& \mS(dt,
\bchi_{ij})^\prime \bdelta_{ij}.
\nonumber
\end{eqnarray}
Identifiability restrictions $f_{ij}(d=0;  \bphi_{ij}, \bbeta_{ij}) =
0$, $g_{ij}(t=0; \bpsi_{ij}, \bgamma_{ij}) = 0$ and $h_{ij}(dt=0;
\bchi_{ij}, \bdelta_{ij}) = 0$ are implemented by fixing $\beta_{ij1} =
0$, $\gamma_{ij1} = 0$ and $\delta_{ij1} = 0$, for all particles and
outcomes (see Figure~\ref{figknots} for an illustration).\vadjust{\goodbreak}

Modeling dose and duration-response curves as piecewise linear
functions allows for considerable flexibility while maintaining direct
interpretability of the model parameters. Recall that in our
formulation the interior knots are estimated as random quantities. This
allows, marginally, for a smooth dose-response trajectory that is
automatically adjusted to fit the data. The main advantage of the
proposed functional representation is that, in the absence of a
dose-time interaction, one can interpret the first interior knot $\phi_{ij1}$ as the dose at which ENM $i$ becomes toxic in relation to
cytotoxicity parameter $j$ (Maximal Safe Dose---similar to the
classical NOAEL concept). A similar interpretation can be given to
$\psi_{ij1}$, in relation to duration-response. Note that the foregoing
interpretation is contingent on fixing $\beta_{ij2} = 0$, $\gamma_{ij2}
= 0$ and $\chi_{ij2} = 0$ when assuming no effect before the first
change-point, and $\beta_{ij2} \leq0$, $\gamma_{ij2} \leq0$ and
$\chi_{ij2} \leq0$ when assuming a tonic effect before the first change
point. In the presence of a dose-time interaction, interpretation
changes slightly and we instead consider the idea of safe exposure
regions, which represent doses and time exposure combinations that do
not induce cytotoxicity. Finally, in the absence of an interaction, the
parameters $\phi_{ij2}$ and $\psi_{ij2}$ are respectively interpreted
as the dose and time at which the response stabilizes, or cells start a
possible recovery process.

We can expand the model further to allow for the exclusion of
interaction functions where not needed.
To do that, we include a latent indicator variable $\rho_{ij}$, so that
for each particle $i$ and outcome $j$
%
\begin{equation}
\label{eqtesting} \qquad%
m_{ij}(d,t) = \cases{ %
\alpha_{ij} + f_{ij}(d; \bphi_{ij},
\bbeta_{ij}) + g_{ij}(t; \bpsi_{ij},
\bgamma_{ij}),  \qquad \mbox{if $\rho_{ij}=0$},\vspace*{2pt}
\cr
\alpha_{ij} + f_{ij}(d; \bphi_{ij},
\bbeta_{ij}) + g_{ij}(t; \bpsi_{ij},
\bgamma_{ij}) + h_{ij}(dt; \bchi_{ij},
\bdelta_{ij}),\vspace*{2pt}\cr
\hspace*{200pt}\mbox{if $\rho_{ij}=1$},}
\end{equation}
where $\rho_{ij} \sim \operatorname{Bern}(\pi)$ and $\pi\sim U(0,1)$. We require that
if $\rho_{ij}=0$, $h_{ij}(dt; \bchi_{ij},\break  \bdelta_{ij})>0$, to ensure
identifiability. The indicator variable $\rho_{ij}$ can then be used to
test explicitly for the dose-time interactions.
The exchangeable Bernoulli trials prior on $\rho_{ij}$ is designed to
account for multiplicities [\citet{Scott2006}].
This trans-dimensional parameterization is key to avoid overfitting, to
facilitate parameter interpretation, and
to allow for testing of specific scientific hypotheses related to the
biological interference of nanomaterials.

For each ENM $i$ and response $j$, we define the following
prior distributions for $\alpha_{ij}$, $\bbeta_{ij}$, $\bgamma_{ij}$,
and $\bdelta_{ij}$:
%
\begin{eqnarray}
\label{eqpriorsslopes} %
\alpha_{ij} &\sim& N\bigl(
\alpha_{o_i},\sigma^{2}_{\alpha_i}\bigr),
\nonumber
\\
\bmm{\beta}_{ij} &\sim& N_4(\bmm{\beta}_{\mathbfo{o}_i},
\bmm{\Sigma}_{\bolds{\beta}
_i}) I \bigl\{ \beta_{ij1} = 0;
\beta_{ij2} \leq0; (\beta_{ij3},\beta_{ij4})\geq 0
\bigr\},
\\
\bmm{\gamma}_{ij} &\sim& N_4(\bmm{
\gamma}_{\mathbfo{o}_i}, \bmm{\Sigma }_{\bolds{\gamma}_i}) I \bigl\{
\gamma_{ij1} = 0; \gamma_{ij2}\leq0; (\gamma_{ij3},
\gamma_{ij4})\geq0 \bigr\},
\nonumber\\
\bmm{\delta}_{ij} \mid\rho_{ij} = 1 &\sim& N_4(
\bmb{m}_{\bolds{\delta}
_{ij}},\bmb {v}_{\bolds{\delta}_{ij}}) I \bigl\{\delta_{ij1} =
0; \delta_{ij2} \leq0; (\delta_{ij3},\delta_{ij4}) > 0
\bigr\}.
\nonumber
\end{eqnarray}
The truncated support for $\bmm{\beta}_{ij}$, $\bmm{\gamma}_{ij}$ and
$\bmm
{\delta}_{ij}$ imposes functional constraints on $f(\cdot)$, $g(\cdot)$
and $h(\cdot)$, which are consistent with the expected behavior of
canonical dose and duration kinetics. At the same time, however, it
allows for the system to recover by permitting a decreasing slope after
the second change point.
The covariance matrix $\bmm{\Sigma}_{\bolds{\beta}_i}$ has diagonal elements
$\bmm
{\sigma}_{\bolds{\beta}_{i\ell}}$, $\ell=1,\ldots,4$, and off diagonal elements
equal to 0; similarly for $\bmm{\Sigma}_{\bolds{\gamma}_i}$.

Prior distributions for $\bphi_{ij}$, $\bpsi_{ij}$ and $\bchi_{ij}$ are
defined to satisfy the following constraints: $(0<\phi_{ij1}<\phi_{ij2}< D)$, $(0 < \psi_{ij1} < \psi_{ij2} < T)$
and $(0<\chi_{ij1}<\chi_{ij2}< DT)$. More precisely, we assume that the joint distribution of
the interior dose and duration knots follows a generalized bivariate
Beta density function, so that
%
\begin{eqnarray}
\label{eqphiknots} %
\phi_{ij} & \sim&B_2(a_{\phi_1},
b_{\phi_1}, a_{\phi_2}, b_{\phi_2}, D),
\nonumber
\\
\psi_{ij} & \sim& B_2(a_{\psi_1}, b_{\psi_1},
a_{\psi_2}, b_{\psi_2}, T),
\\
\chi_{ij} & \sim& B_2(a_{\chi_1}, b_{\chi_1},
a_{\chi_2}, b_{\chi
_2}, DT).
\nonumber
\end{eqnarray}
Here we assume that a random vector $\mathbf{x} = (x_1, x_2)'$ is
distributed according to a generalized bivariate
Beta distribution function ($\mathbf{x} \sim B_2(a_1, b_1, a_2, b_2, m)$),
with support $\mathcal{S}(\mathbf{x}) = \{ (x_1, x_2)\dvtx  0 < x_1 < x_2 <
m\}
$ if and only if
%
\begin{eqnarray}
\label{eqgenbeta} %
&& p( \mathbf{x} \mid a_1,
b_1, a_2, b_2, m)
\nonumber
\\
&&\qquad= p(x_1 \mid a_1, b_1, m)
p(x_2 \mid x_1, a_2, b_2, m)
\\
&&\qquad= \frac{1}{B(a_1, b_1)} \frac{x_1^{a_1 - 1} (m -
x_1)^{b_1 - 1} }{m^{a_1 + b_1 - 1}} \frac{1}{B(a_2, b_2)}
\frac{ (x_2 - x_1)^{a_2 - 1} (m -
x_2)^{b_2 - 1} }{(m - x_1)^{a_2 + b_2 - 1}}.
\nonumber
\end{eqnarray}
The foregoing formulation can be seen as a generalization of the
Dirichlet distribution over
a two-dimensional simplex. This general formulation can be simplified
further, in order to achieve a right-skewed marginal distribution for $x_1$
and a uniform conditional distribution for $x_2$ given $x_1$. This is
achieved by assuming $b_1 > a_1 > 1$ and $a_2 = b_2 = 1$.

Making use of this construction, we simplify the prior distribution in
(\ref{eqphiknots}) as follows:
%
\begin{eqnarray}
\label{eqphiknotsreduced} %
\phi_{ij} & \sim&B_2(1,
\lambda_{\phi_{i 1}}, \lambda_{\phi_{i
2}}, 1, 1, D) I\{\lambda_{\phi_{i 2}}
> \lambda_{\phi_{i 1}} > 1 \},
\nonumber
\\
\psi_{ij} & \sim& B_2(1, \lambda_{\psi_{i 1}},
\lambda_{\psi_{i 2}}, 1, 1, T) I\{\lambda_{\psi_{i 2}} > \lambda_{\psi_{i 1}}
> 1 \},
\\
\chi_{ij} & \sim& B_2(1, l_{\chi_{i 1}},
l_{\chi_{i 2}}, 1, 1, T) I\{ l_{\chi_{i 2}} > l_{\chi_{i 1}} > 1 \}.
\nonumber
\end{eqnarray}
From a regulatory standpoint, this formulation favors (a priori) the
choice of conservative values for the location of the first change
point and a relatively diffuse prior distribution for our second change
point (see Figure~\ref{figknots}).

\textit{Stage \textup{3:} Response model at the ENM level.}
For each ENM $i$, we exploit conditional conjugacy to define the
following prior distributions for population level parameters:
%
\begin{equation}
\label{eqstage3} 
\qquad\alpha_{o_i} \sim N(m_{\alpha_i},v_{\alpha_i}),
\qquad \bmm{\beta}_{\mathbfo{o}_i} \sim N_4(\bmb{m}_{\bolds{\beta}_i},
\bmb{v}_{\bolds{\beta}_i}),\qquad \bmm{\gamma}_{\mathbfo{o}_i} \sim N_4(
\bmb{m}_{\bolds{\gamma}_i}, \bmb{v}_{\bolds{\gamma}_i}). 
\end{equation}
In the absence of an interaction, the parameters $\bmm{\beta}_{\mathbfo{o}_i}$ and
$\bmm{\gamma}_{\mathbfo{o}_i}$ represent summaries of the dose and
duration-response trajectories across all outcomes and the $\alpha_{o_i}$ parameters represent a summary of the baseline response across
all outcomes.
In the presence of an interaction, we may construct these summaries
conditionally on specific doses and durations of exposure.

Finally, considering the distribution introduced in (\ref
{eqphiknotsreduced}), we define a prior model for population level
parameters $\bmm{\lambda}_{\bolds{\phi}_i}=(\lambda_{\phi_{i 1}}, \lambda_{\phi
_{i 2}})$ and $\bmm{\lambda}_{\bolds{\psi}_i}=(\lambda_{\psi_{i 1}},
\lambda_{\psi_{i 2}})$ as follows:
%
\begin{equation}
\label{eqknotspop} %
\lambda_{\phi_{i \ell}} \sim \operatorname{Gamma}(a_{\lambda_{\phi i \ell
}},b_{\lambda
_{\phi i \ell}}),
\qquad \lambda_{\psi_{i \ell}} \sim \operatorname{Gamma}(a_{\lambda_{\psi i \ell
}},b_{\lambda
_{\psi i \ell}}),
\end{equation}
where $\ell= 1, 2$. The parameters $\bmm{\lambda}_{\bolds{\phi}_i}$ and $\bmm
{\lambda}_{\bolds{\psi}_i}$ can be used to construct summaries of dose and
duration-response change points across all outcomes.
Shape hyperparameters ($a_{\lambda_{\phi i\ell}},b_{\lambda_{\phi
i\ell
}}$) and ($a_{\lambda_{\psi i\ell}},b_{\lambda_{\psi i\ell}}$)
can be tuned to favor more or less conservative values for the
change-point locations at the particle level.


\textit{Stage \textup{4:} Hyperpriors}.
We complete the model by specifying prior distributions on our
hyperparameters as follows:
%
\begin{eqnarray}
\label{eqstage4} %
1/\sigma^{2}_{\varepsilon_j} &\sim&
\operatorname{Gamma}(a_{\varepsilon_j},b_{\varepsilon_j}), \qquad 1/
\sigma^{2}_{\alpha_i} \sim \operatorname{Gamma}(a_{\alpha_i},b_{\alpha_i}),
\nonumber
\\[-8pt]
\\[-8pt]
\nonumber
1/\sigma^{2}_{\beta_i} &\sim& \operatorname{Gamma}(a_{\beta_i},b_{\beta_i}),
\qquad 1/\sigma^{2}_{\gamma_i} \sim \operatorname{Gamma}(a_{\gamma_i},b_{\gamma_i}).
\end{eqnarray}
We model our precision parameters as gamma distributions, exploiting
conditional conjugacy.
Again, prior parameters can be tuned to define more or less informative
distributions consistent
with the scale of the outcomes [\citet{Gelman2006}]. Note that in our
formulation, $x \sim \operatorname{Gamma}(a,b)$ denotes a $\operatorname{Gamma}$ distributed random
quantity with shape $a$ and rate $b$, such that $E(x) = a/b$.

\section{Estimation and inference} \label{subsecestinf}
\subsection{Posterior simulation via MCMC} \label{subsubsectionest}

Using the B-spline representation introduced in Section~\ref
{subsecDescription}, we can write the expected $j$th response level
associated with
ENM $i$, at dose $d$ and exposure time $t$ as
\[
m_{ij}(d,t; \alpha_{ij}, \bbeta_{ij},
\ldots) = \cases{ %
\alpha_{ij} + \mS(d,
\bphi_{ij})'\bbeta_{ij} + \mS(t,
\bpsi_{ij})'\bgamma_{ij},  \qquad\mbox{if $\rho_{ij}=0$},\vspace*{2pt}
\cr
\alpha_{ij} + \mS(d,
\bphi_{ij})'\bbeta_{ij} + \mS(t,
\bpsi_{ij})'\bgamma_{ij} + \mS(dt,
\bchi_{ij})'\bdelta_{ij}, \vspace*{2pt}\cr
\hspace*{184pt}\mbox{if $
\rho_{ij}=1$}.}
\]

Let $\bbeta= \{\bbeta_{ij}\dvtx    i=1,\ldots,I,  j=1,\ldots,J\}$ and
define $\bgamma$ and $\bdelta$ in a similar fashion.
These parameters denote the full set of spline coefficients.
Furthermore, consider knot parameters $\bphi= \{\bphi_{ij}\dvtx
i=1,\ldots,I, j=1,\ldots,J\}$, with $\bpsi$
and $\bchi$ similarly defined, and background response parameters
$\bolds{\alpha} = \{\alpha_{ij}\dvtx    i=1,\ldots,I, j=1,\ldots
,J\}$.
Finally, let $\mbfs{\sigma}^2_\varepsilon= (\sigma^2_{\varepsilon
_1},\ldots
,\sigma^2_{\varepsilon_J})'$ and $\mbfs{\tau} = (\tau_1, \ldots,
\tau_I)'$. If we denote with $\mathbf{Y}$ the complete set of response values
for all particles and cytotoxicity outcomes, the likelihood function
can be written as follows:
%
\begin{eqnarray}
\label{eqlikelihood}&& L\bigl( \bbeta, \bgamma,\bdelta, \bphi, \bpsi, \bchi,
\bolds{\alpha}, \mbfs{\sigma}^2_\varepsilon, \mbfs{\tau}, \mbfs{\rho} \mid\mathbf{Y}\bigr)
\nonumber
\\[-8pt]
\\[-8pt]
\nonumber
&&\qquad\propto\prod_{i,j,k,d,t} \biggl[
\biggl(\frac{\sigma^2_{\varepsilon
_j}}{\tau_i} \biggr)^{-{1}/{2}} \exp \biggl\{-\frac{( y_{ijk}(d,t) -
m_{ij}(d,t;\ldots) )^2}{2\sigma^2_{\varepsilon_j} / \tau_i} \biggr\} \biggr],
\end{eqnarray}
where the product is taken over all replicates $k$, particles $i$,
outcomes $j$, doses $d$ and times $t$. We are interested in the
posterior distribution
%
\begin{eqnarray}
\label{eqPosterior} %
\qquad P\bigl(\bbeta, \bgamma,
\bdelta, \bphi, \bpsi, \bchi, \bolds{\alpha}, \mbfs{\sigma}^2_\varepsilon,
\mbfs{\tau}, \mbfs{\rho} \mid\mathbf{Y}\bigr)& \propto&L\bigl(\bbeta, \bgamma,
\bdelta, \bphi, \bpsi, \bchi, \bolds{\alpha}, \mbfs{\sigma}^2_\varepsilon,
\mbfs{\tau}, \mbfs{\rho} \mid{\bf Y}\bigr)
\nonumber
\\[-8pt]
\\[-8pt]
\nonumber
&&{}\times P\bigl(\bbeta, \bgamma, \bdelta, \bphi, \bpsi, \bchi, \bolds{\alpha},
\mbfs{\sigma}^2_\varepsilon, \mbfs{\tau}, \mbfs{\rho} \bigr),
\end{eqnarray}
where the prior model $P(\bbeta, \bgamma, \bdelta, \bphi, \bpsi,
\bchi,
\bolds{\alpha}, \mbfs{\sigma}^2_\varepsilon, \mbfs{\tau},
\mbfs{\rho
})$ is fully described in Section~\ref{subsecDescription}.
This quantity is, however, unavailable in closed analytic form,
therefore, we base our inference on Markov Chain Monte Carlo (MCMC)
simulations.

The proposed posterior simulation algorithm combines Gibbs
steps within Metropolis--Hastings steps in a hybrid sampler, where we
update parameters component-wise [\citet{Tierney1994}]. We directly
sample components when closed-form full conditional distributions are
available using a Gibbs sampling algorithm [\citet{Geman1984}, \citet
{Gelfand1990}]; otherwise, we use the Metropolis--Hastings (MH) approach
[Metropolis et al. (1953)]. Available full conditional distributions
are given in the supplemental article, Appendix A [\citet{Patel2012}].
As we are considering selection of interaction functions in a
trans-dimensional setting, we implement a reversible jumps algorithm to
move between models with and without the dose-time interaction function
$h_{ij}(dt; \bchi_{ij}, \bdelta_{ij})$ [\citet{Green1995}]. The model
indicator $\rho_{ij}$ and corresponding model parameters $\bdelta_{ij}$
and $\bchi_{ij}$ are updated jointly using reversible jump MCMC steps.
After the model structure has been specified, the model parameters are
updated from their corresponding conditional posterior distributions.
The proposed sampling scheme can be summarized as follows.

1. \textit{Fixed dimensional updates.}
Given the current state of the latent interaction indicators $\rho_{ij}$,
response surfaces are uniquely defined as in (\ref{eqtesting}).
Posterior sampling is standard here and proceeds by updating spline
coefficients $\bbeta,\bgamma$ and $\bdelta$
from their conditional posterior via direct simulation [\citet
{Patel2012}]. Knot parameters $\bphi, \bpsi$ and $\bchi$
are updated via a MH step.
For example, when sampling the interior knot parameters $\bmm{\phi}$
we use an appropriate proposal kernel $q(\phi_{ij\ell}^{0},\phi_{ij\ell}^{1})$
to efficiently construct Markov chains with the desired stationary
distribution. While accounting for the fact that $\phi_{ij1}<\phi_{ij2}$,
we consider uniform proposal densities of the form
%
\begin{equation}
\label{eqproposal} %
q\bigl(\phi_{ij\ell}^{1}
\mid\phi_{ij\ell}^{0}\bigr) = U\bigl(\phi_{ij\ell
}^{0}-w_{\phi ij\ell},
\phi_{ij\ell}^{0}+w_{\phi ij\ell}\bigr)I(S_{\phi
}),
\end{equation}
where $\ell=1,2$. Here $S_{\phi}$
denotes the appropriate support and must satisfy the constraints
$0<\phi_{ij1}<\phi_{ij2}<D$.
Proposed values of $\phi_{ij\ell}$
are accepted with the following probabilities:

\begin{equation}
\label{eqaratio} %
\min \biggl\{1;
\frac{p(\phi_{ij\ell}^{1} \mid
y_{ijk}, \bmm
{\theta}_{\setminus\phi})}{p(\phi_{ij\ell}^{0} \mid y_{ijk}, \bmm
{\theta
}_{\setminus\phi})} \frac{q(\phi_{ij\ell}^{0} \mid\phi_{ij\ell
}^{1})}{q(\phi_{ij\ell}^{1} \mid\phi_{ij\ell}^{0})} \biggr\},\qquad \ell=1,2.
\end{equation}
To tune proposal kernels, each $\phi_{ij\ell}$
was sampled using an initial value of $w$ that was re-calibrated
throughout the burn-in period to achieve an acceptance rate between
$30\%$ and $70\%$ [\citet{Roberts2001}]. Specifically, the acceptance
rate of $\phi_{ij\ell}$
was monitored every 200 iterations throughout the burn-in period with
$w_{\phi ij\ell}$
adjusted appropriately if the acceptance rate did not fall within the
desired range.
A similar Metropolis--Hastings scheme was adapted for sampling the
duration-response parameters $\bpsi$, dose-time interaction parameters
$\bchi_{ij}\mid\rho_{ij}=1$, as well as for population level knot parameters.

2. \textit{Trans-dimensional updates.} We sample the
model space by randomly proposing
the birth or death of dose-time interaction functions $h_{ij}(\cdot)$.
This is accomplished by
selecting a particle $i$ and outcome $j$ at random and by jointly
updating $\rho_{ij}$, $\bdelta_{ij}$ and $\bchi_{ij}$.
In detail,

\begin{longlist}[(1)]
\item[(1)] For uniformly random $i \in(1,\ldots, I)$ and $j \in(1,\ldots, J)$,
propose a systematic change $\rho_{ij}^{0} \rightarrow\rho_{ij}^{1}=1-\rho_{ij}^{0}$.
We assume for the moment that we propose moving from $\rho_{ij}^0 = 0$
to $\rho^1_{ij} = 1$, implying the birth of a new interaction function
$h_{ij}(\cdot)$.
\item[(2)] Propose new knots and spline coefficients $\bdelta_{ij}^{1}\sim
q(\bdelta_{ij}^{1})$ and $\bchi_{ij}^{1} \sim q(\bchi_{ij}^{1})$.
\item[(3)] Accept the proposed move with probability $\tau_b=\min(1, R_b)$, where
%
\begin{equation}
\label{eqtrans} R_b=\frac{p(y_{ijk} \mid\bdelta_{ij}^{1}, \bchi_{ij}^{1}, \rho_{ij}^{1}, \bmm{\theta}_{\setminus{\bmm{\delta}_{ij}, \bmm{\chi}_{ij},
\rho_{ij} }}) } {
p(y_{ijk} \mid\rho_{ij}^{0}, \bmm{\theta}_{\setminus{\bmm{\delta
}_{ij}, \bmm{\chi}_{ij}, \rho_{ij} } })} \frac{p(\bdelta_{ij}^{1} \mid\rho_{ij}^{1})p(\bchi_{ij}^{1} \mid
\rho_{ij}^{1}) }{q(\bdelta_{ij}^1) q(\bchi_{ij}^{1})}
\frac{p(\rho_{ij}^1)}{p(\rho_{ij}^0)},
\end{equation}
where we use $\bmm{\theta}_{\setminus\omega}$ to denote all model
parameters, with the exception of $\omega$.
\end{longlist}
In the case where the proposed move would imply a death of an
interaction function ($\rho_{ij}^0 = 1 \rightarrow\rho_{ij}^1=0$), the
acceptance probability
would simply be $\tau_d = 1/\tau_b$.

While the proposal densities $q(\bdelta_{ij})$ $q(\bchi_{ij})$ in
(\ref
{eqtrans}) can in theory be defined almost arbitrarily, to guarantee
efficient exploration of the
model space, we consider truncated multivariate normal proposals for
$\bdelta_{ij}$ and $\bchi_{ij}$ centered around regions of high
posterior probability.
Efficient optimization within the MCMC iterations is achieved using
standard profile likelihood ideas [\citet{Severini1994}].

\subsection{Posterior inference} \label{subsubsectioninf}
In this section we discuss inference on ENM-specific risk assessment
parameters, based on draws from
the posterior distribution described in Section~\ref{subsubsectionest}. Table~\ref{tabriskpar} summarizes several quantities of interest
including the maximal safe dose, maximal safe exposure time and the
maximal response. This list is not exhaustive. However, other risk
assessment parameters of interest, such as benchmark doses (BMD) or
effective concentrations (EC$\alpha$), are easily obtained from our
model output in a numerical fashion. In the case of a dose-time
interaction, these quantities are defined conditionally on specific
doses and durations of exposure.

%
\begin{table}
\caption{Risk assessment
parameters. ENM level risk assessment parameters associated with the
hierarchical model introduced in \protect\ref{subsecDescription}. For each
parameter we summarize its function in the model and the related
interpretation as a cytotoxicity risk factor}\label{tabriskpar}
\begin{tabular*}{\textwidth}{@{\extracolsep{\fill}}lcc@{}}
\hline
\multicolumn{1}{@{}l}{\textbf{Parameter}} & \textbf{Model function} & \multicolumn{1}{c@{}}{\textbf{Parameter interpretation}} \\
\hline
$\beta_{3ij}^{*}$ & Dose-response slope from $\phi_{ij1}$ to $\phi_{ij2}$ & Overall dose effect \\
$\gamma_{3ij}^{*}$ & Duration-response slope from $\phi_{ij1}$ to
$\phi_{ij2}$ & Overall exposure time effect \\
$\phi_{1ij}$ & Dose-response change point 1 & Maximal safe dose \\
$\psi_{1ij}$ & Duration-response change point 1 & Maximal safe
exposure time \\
$m_{ij}^*$ & Evaluated numerically & Maximal response \\
\hline
\end{tabular*}
\end{table}

\begin{figure}

\includegraphics{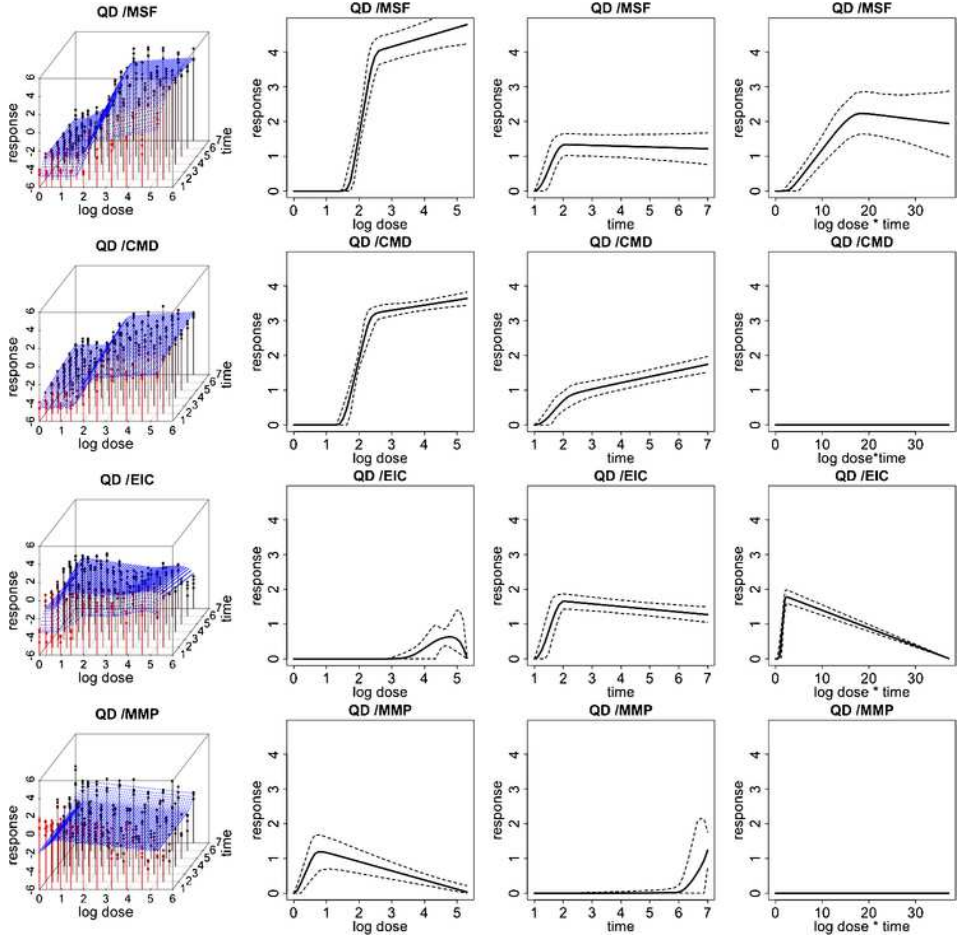}

\caption{Fitted
response curves for the quantum dot (QD) ENM. (\textit{left}) Fitted response
surfaces (\textit{column} 1), dose-response function, $f_{ij}(d)$
(\textit{column} 2),
duration-response function, $g_{ij}(t)$ (\textit{column}~3),
dose/duration interaction
function, $h_{ij}(dt)$ (\textit{column} 4) and associated 95\% posterior
intervals.
In (\textit{column} 1), the color red represents response values
corresponding to lower
time points and the color black represents response values
corresponding to higher time points.}
\label{figfittedQD}
\end{figure}

\begin{figure}

\includegraphics{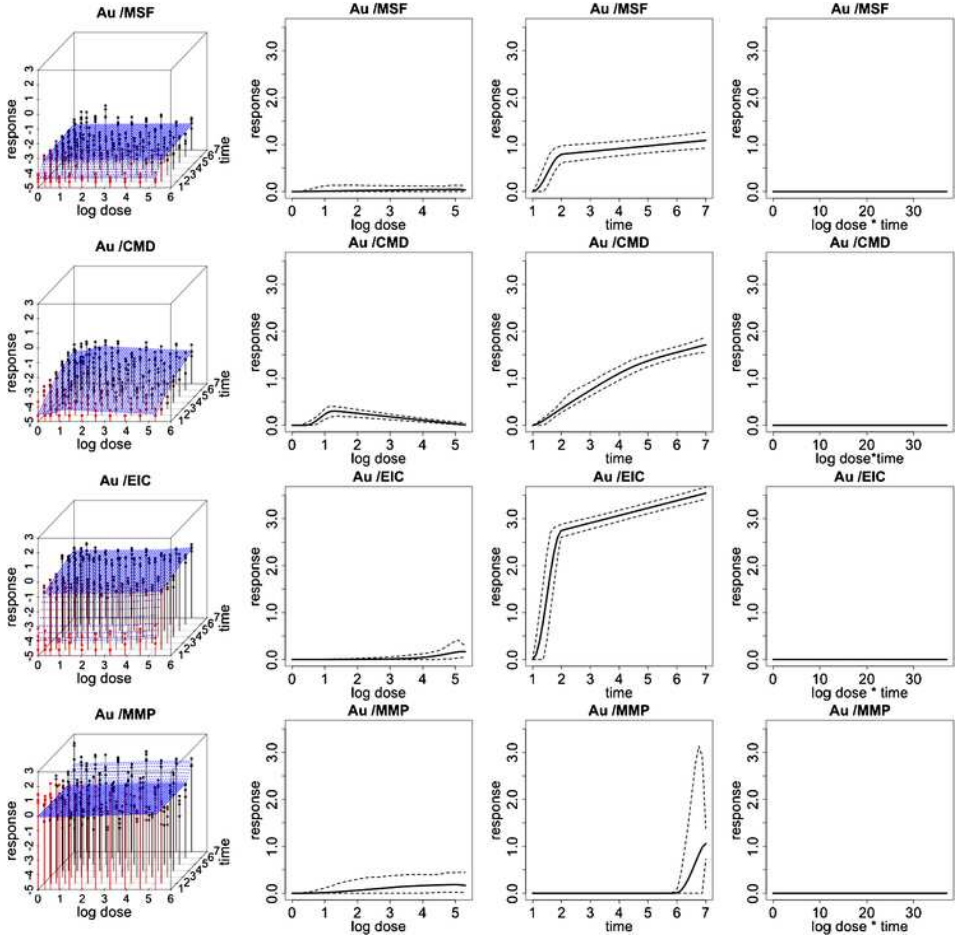}

\caption{Fitted response curves for the gold (Au) ENM.
Fitted response surfaces (\textit{column}~1), dose-response function,
$f_{ij}(d)$ (\textit{column}~2),
duration-response function, $g_{ij}(t)$ (\textit{column}~3),
dose/duration interaction function, $h_{ij}(dt)$
(\textit{column} 4) and associated 95\% posterior intervals. In
(\textit{column} 1), the color red represents
response values corresponding to lower time points and the color black
represents response values
corresponding to higher time points.}
\label{figfittedau}
\end{figure}

Let $\bmm{\phi}_{ij}^{(n)}$, $\bmm{\psi}_{ij}^{(n)}$, $\bmm
{\chi}_{ij}^{(n)}$, $\bmm{\beta}_{ij}^{(n)}$, $\bmm{\gamma}_{ij}^{(n)}$,
$\bmm
{\delta}_{ij}^{(n)}$, $\alpha_{ij}^{(n)}$ and $\rho_{ij}^{(n)}$,
$n=1,\ldots,N$, denote~$N$ MCMC draws from the posterior distribution of
$\bmm{\phi}_{ij}$, $\bmm{\psi}_{ij}$, $\bmm{\chi}_{ij}$, $\bmm{\beta
}_{ij}$, $\bmm{\gamma}_{ij}$, $\alpha_{ij}$ and~$\rho_{ij}$. In the
absence of an interaction term, posterior samples $\phi_{ij1}^{(n)}$
and $\psi_{ij1}^{(n)}$ directly provide us with an approximation of the
posterior distribution for the maximal safe dose and maximal safe
exposure time. We can also obtain the posterior samples for the overall
dose effect, $\beta_{ij3}^{*(n)} = \beta_{ij3}^{(n)}/(\phi_{ij2}^{(n)}-\phi_{ij1}^{(n)})$, which is the slope of the
dose-response curve between $\phi_{ij1}$ and $\phi_{ij2}$. Similarly,
we can obtain the posterior distribution for the overall time effect
using posterior samples $\gamma_{ij3}^{*(n)} = \gamma_{ij3}^{(n)}/(\psi_{ij2}^{(n)}-\psi_{ij1}^{(n)})$.
In the presence of a dose-time
interaction, we can define any of the summaries described above
conditionally on a given dose and time. For example, the maximal safe
dose conditional on exposure time can be defined as $\min\{\phi_{ij1},
\chi_{ij1} /t\}$, and posterior samples can be obtained from $\min\{
\phi_{ij1}^{(n)}, \chi_{ij1}^{(n)} /t\}$. Given posterior draws, one can
proceed with the straightforward construction of standard posterior
summaries, such as means, maxima a posteriori, modes, quantiles and
credible regions. We may also be interested in testing for a dose-time
interaction. The expected inclusion probability of the dose-time
interaction function can be estimated using posterior draws $\rho_{ij}^{(n)}$ as
$\hat{p}_{ij} = \sum_{n} \rho_{ij}^{(n)} /N$. Given the
prior distribution described in (\ref{eqtesting}), this posterior
probability is known to adjust for multiplicities and can be used to
test for a dose-time interaction. \citet{Scott2006}, for example,
recommend selecting the median model, that is, including all
interactions for which $\hat{p}_{ij} > 0.5$. Also of interest is an
estimate of the dose-response surface, $m_{ij}(d,t)$, for particle $i$
and outcome $j$. This surface is, of course, defined in an
infinite-dimensional space. However, given the basis-function
representation introduced in Section~\ref{subsecDescription}, we only
need finite draws from the parameter set of interest. More precisely,
draws from the marginal posterior distribution of the dose-response
surface for any dose $d \in[0,D]$ and time $t\in[0,T]$ are given by
%
\begin{equation}
\label{eqsurface}
m_{ij}^{(n)}(d,t)
= \cases{
\alpha_{ij}^{(n)}
+ \mS\bigl(d, \bphi_{ij}^{(n)}\bigr)'
\bbeta_{ij}^{(n)} + \mS \bigl(t, \bpsi_{ij}^{(n)}
\bigr)'\bgamma_{ij}^{(n)} ,\qquad \mbox{if $
\rho_{ij}^{(n)}=0$},
\vspace*{2pt}\cr
\alpha_{ij}^{(n)} + \mS\bigl(d, \bphi_{ij}^{(n)}
\bigr)'\bbeta_{ij}^{(n)} + \mS \bigl(t,
\bpsi_{ij}^{(n)}\bigr)'\bgamma_{ij}^{(n)}
+ \mS\bigl(dt, \bchi_{ij}^{(n)}\bigr)'
\bdelta_{ij}^{(n)},\vspace*{2pt}\cr
\hspace*{204pt}\mbox{if $\rho_{ij}^{(n)}=1$}.}\hspace*{-28pt}
\end{equation}
For each $\bmm{\phi}_{ij}^{(n)}$, $\bmm{\psi}_{ij}^{(n)}$, $\bmm
{\beta}_{ij}^{(n)}$, $\bmm{\phi}_{ij}^{(n)}$ and $\alpha_{ij}^{(n)}$,
$n=1,\ldots,N$, we evaluate the dose-response function given in (\ref
{eqsurface}) over a grid of values $\tilde{D} = (d_1,\ldots,d_n)'$ and
$\tilde{T} = (t_1, \ldots, t_n)'$. The posterior mean of the samples
$m_{ij}^{(n)}$, $n=1,\ldots,N$, at each value of $\tilde{D}$ and $\tilde
{T}$ can be used to summarize the fit of the dose-response surface, as
shown in Figures~\ref{figfittedQD} to~\ref{figfittedau}. Other
quantities of interest include the posterior distribution of the
dose-response function $f_{ij}(d; \bphi_{ij},\bbeta_{ij})$,
duration-response function $g_{ij}(t; \bpsi_{ij},\bgamma_{ij})$ and
dose-time interaction function $h_{ij}(dt; \chi_{ij},\bdelta_{ij})$.
Draws from the marginal posterior distribution of these functions for
any dose $d \in[0, D]$ and time $t \in[0, T]$ are given by
%
\begin{eqnarray}
\label{eqDDfxns} %
f_{ij}^{(n)}(d;
\bphi_{ij},\bbeta_{ij}) &=& \mS\bigl(d,\phi_{ij}^{(n)}
\bigr)'\bbeta_{ij}^{(n)},
\nonumber\\
g_{ij}^{(n)}(t; \bpsi_{ij},\bgamma_{ij})
&=& \mS\bigl(t,\psi_{ij}^{(n)}\bigr)'
\bgamma_{ij}^{(n)},
\\
h_{ij}^{(n)}(dt; \bchi_{ij},\bdelta_{ij})
&=& \mS\bigl(dt,\chi_{ij}^{(n)}\bigr)'
\bdelta_{ij}^{(n)}.\nonumber
\end{eqnarray}
For each draw, we evaluate the dose-response functions over a grid of
values $d\in\tilde{D}$ and the duration-response functions over a grid
of values $t\in\tilde{T}$.
As described before, standard pointwise posterior summaries can be
obtained in a straightforward fashion. Simultaneous confidence bands
for the functional effect of interest can be constructed following the
Monte Carlo approximation suggested by \citet{Veera2005}.

Additional summaries of interest can be obtained in a numerical
fashion. For example, the posterior distribution for the maximal
response value $m_{ij}^* = \max\{m_{ij}(d,t); d\in[0, D], t\in
[0,T]\}
$, may be obtained evaluating $m_{ij}^{(n)}(d,t)$ over a fine grid of
doses $\tilde{D} $ and times $\tilde{T} $. An approximate posterior
draw from $m^*_{ij}$ can be defined as $m^{*(n)}_{ij} = \max\{
m^{(n)}_{ij}(d,t); d\in\tilde{D}, t\in\tilde{T}\}$. Given
smoothness constraints on $m_{ij}(d,t)$, defined in Section~\ref
{subsecDescription}, the foregoing procedure is likely to provide
a good approximation to the posterior distribution of the maximal
response value, provided $\tilde{D}$ and $\tilde{T}$ define
a sufficiently detailed evaluation grid. Similar procedures may be
adopted to
obtain inference on other risk assessment parameters like EC$\alpha$s
or BMDs.

\section{Applications}
\label{secApplication}

\subsection{Synthetic data}\label{susecSimulation}

To assess estimation of the model presented in Section~\ref
{secModel}, we present a simulation study in the supplemental article,
Appendix B [\citet{Patel2012}]. The dose and time kinetics were
simulated in an additive fashion, from various parametric functions,
including both canonical and noncanonical profiles that are still
reasonably interpretable under a toxicity framework. We also placed
increasingly conservative priors on the population level parameters
$\bmm
{\lambda}_{\bolds{\phi}_i}$ and $\bmm{\lambda}_{\bolds{\phi}_i}$ in order to assess the
sensitivity of the model results to our choice of prior parameters. In
the supplemental article, Appendix C, we provide an additional
sensitivity analysis assessing model results to our choice of prior
model for the change-point parameters [\citet{Patel2012}]. We compare
our prior model results to both a truncated normal prior and a
parameterization of the bivariate beta prior that results in a uniform
prior on the simplex.

Simulation results indicate that our model is robust to model
misspecification and is not very sensitive to our choice of prior. We
do, however, maintain that using the bivariate beta prior defined in
(8) is likely to be more appropriate in data analytic frameworks, as
the implied stochastic behavior of the response surface, {a
priori}, reflects more closely the usual biological mechanisms of
toxicity. More specifically, it assigns zero probability of toxicity to
zero dose and time, where toxicity is indeed not expected to occur.
Furthermore, this prior accounts for issues such as dosimetry, in which
the administered doses are confounded by different particle
bioavailability. Therefore, in some particles toxicity is not expected
to occur for doses and times greater than zero.

\subsection{Case study background} \label{subsecBackground}
We illustrate the proposed methodology by analyzing data on
macrophage cells (RAW cells) exposed to eight different metal and
metal-oxide nanoparticles, monitored in relation to four cytotoxicity
parameters. All four outcomes are measured over a grid of ten doses and
seven times (hours) of exposure (see Figures~\ref{figfittedQD} to \ref
{figfittedau}). Cytotoxicity screening is based on the hierarchical
oxidative stress model [\citet{GeorgePokhrel2010}]. More specifically, a
multi-parametric assay that utilizes four compatible dye combinations
and the subsequent change in fluorescence read-outs was used to measure
four responses relating to the highest tier of oxidative stress (toxic
oxidative stress). The four measured responses include mitochondrial
superoxide formation (MSF), loss of mitochondrial membrane potential
(MMP), elevated intracellular calcium (EIC) and cellular membrane
damage (CMD).
Figure~\ref{figraw} provides fluorescence images of cells exposed to
various nanomaterials (50 $\mu$g/ml and 3 hours), including quantum
dot, platinum and a negative control consisting of no nanomaterials.
\textit{Row} 1 includes images of cells treated with a dye combination
including MitoSox, which permeates the mitochondria and fluoresces red
when oxidized by superoxide. Red fluorescence measured in cells treated
with MitoSox is therefore a measure of mitochondrial superoxide
formation. Similarly, in \textit{Row} 2 cells are treated with a dye
combination including JC1, which stains the cytoplasm red in healthy
cells, but forms a monomer in cells with decreased membrane potential
and consequently stains the cytoplasm green. Finally, in \textit{Row} 3
cells are strained with a dye combination including Fluo-4 and
Propidium Iodide (PI). In cells with damaged membranes, PI is able to
permeate the cell and bind to DNA where it causes the nucleus to emit a
red florescence. Fluo-4 is a dye that emits a green fluorescence in the
cytoplasm in cells with elevated intracellular calcium. Each sample was
also stained with a Hoechst dye which causes all cell nuclei to emit a
blue florescence, allowing for a count of the total number of cells. An
analysis of the fluorescence readout, monitored at varying wavelengths,
results in a measure of the percentage of cells positive for each
response. Figure~\ref{figraw} also provides a heat map of the raw
responses for each particle and outcome, where colder colors (blues and
greens) indicate a smaller percentages of cells positive for the
response and warmer colors (oranges and reds) indicate a higher
percentage of cells positive for the response. The final data was
normalized using a logit transformation to unconstrain the support so
that it can take on values between $-\infty$ and $\infty$.
Our inferences are based on 20,000 MCMC samples from the posterior
distribution in (\ref{eqPosterior}), after discarding a conservative
60,000 iterations for burn-in. MCMC sampling was performed in R version~2.10.0, and convergence diagnostics were performed using the package
CODA (Convergence Diagnostics and Output Analysis), [\citet{coda2006}].

\begin{figure}

\includegraphics{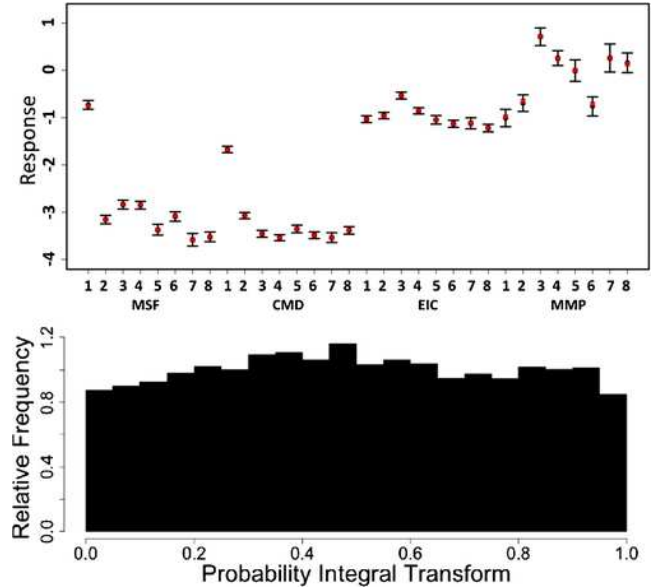}

\caption{Graphical model
diagnostics. (\textit{Bottom}) Probability Integral Transform assessing
empirical calibration of the posterior predictive distribution.
(\textit{Top}) Mean and 95\% posterior intervals of the posterior predictive
mean response across all doses and times of exposure, for all outcomes
and particles 1 through 8 (QD, ZnO, Fe$_3$O$_4$, Pt, Ag, SiO$_2$,
Al$_2$O$_3$, Au). Also included are the empirical mean responses across
all doses and times of exposure (red).}
\label{figdiagnostics2}
\end{figure}

\subsection{Case study analysis and results} \label{subsecAnalysis}
We fit the model described in Section~\ref{subsecDescription} to the
metal-oxide data set described in the previous section.
The prior on the interior knot parameters was modeled using the
simplified density described in~(\ref{eqphiknotsreduced}). A set of
relatively noninformative $\operatorname{Gamma}(2,1)$ and $\operatorname{Gamma}(3,1)$ priors were
considered for the components of both $\bmm{\lambda}_{\bolds{\phi}_i}$ and
$\bmm
{\lambda}_{\bolds{\psi}_i}$, along with a vague $B_2(2,3,1,1, DT)$ prior for
our dose-time interaction change-point parameter $\bchi_{ij}$.
We also fixed $\beta_{ij2} = 0$ and $\gamma_{ij2} = 0$, assuming no
effect before $\phi_{ij1}$ and $\psi_{ij1}$, thereby allowing, in the
absence of a dose-time interaction, the interpretation of $\phi_{ij1}$
as the maximal safe dose and $\psi_{ij1}$ as the maximal safe exposure
time. Similarly, when $\rho_{ij}=1$, we fixed $\delta_{ij2} = 0$.
We placed $\operatorname{Gamma}(0.01,0.01)$ priors on the $1/\sigma_{\varepsilon_{j}}$
parameters, $\operatorname{Gamma}(1,0.1)$ priors on all remaining precision parameters
and $N(0,10)$ priors on the $\alpha_{oi}$ parameters. The parameters
$\bmm{\beta}_{{o}i}$ and $\bmm{\gamma}_{{o}i}$ are modeled as truncated
multivariate normals with mean $\bmb{1}$ and a covariance matrix with
diagonal elements 10 and off-diagonal elements 0. Finally, we placed a
prior distribution on the degrees of freedom parameter $\nu$, for the
$T$-distributed error described in Section~\ref{subsecDescription}. We
specified the prior to be uniform on 1, 2, 4, 8, 16 and 32 degrees of
freedom [\citet{Besag1999}].
In concordance with our synthetic data experiments, a sensitivity
analysis on the case study data set proved robust to reasonable
variations in the prior specification.

We provide graphical summaries of goodness of fit and
posterior predictive performance in Figure~\ref{figdiagnostics2}.
The top panel shows the mean and 95\% posterior intervals of the
posterior predictive mean response across all doses and times of
exposure (black), along with the empirical mean response (red), for
each particle and outcome. In all cases the empirical mean response is
contained within the 95\% posterior intervals of the posterior
predictive mean distribution, indicating good average posterior
coverage across doses and times of exposure.
The bottom panel provides a plot of the probability integral transform
histogram for the entire model [\citet{Gneiting2007}]. Visual assessment
of the plot indicates that it is close to uniformity, suggesting
relatively good posterior predictive calibration.
Additional summaries and diagnostic tools are detailed in the
supplemental article, Appendix E [\citet{Patel2012}].

Figures~\ref{figfittedQD} and~\ref{figfittedau} illustrate data and
results associated with two of the particles examined in this HTS
study. Particularly, we report inference for platinum and quantum dot
nanomaterials for each of the 4 cytotoxicity outcomes. Inference for
the remaining 6 particles is reported in the supplemental article,
Appendix~D [\citet{Patel2012}]. In these two figures, \textit{column} 1
shows expected posterior dose-response surfaces across dose and time
for all outcomes. As the posterior expectation marginalizes over the
interior knots, smooth surfaces reflect the uncertainty about the
location of these change points and provide an illustration of
how the proposed technique will adjust for smoothness in an
unsupervised fashion. Also included are functional posterior
expectations associated with dose-response curves $f_{ij}(d)$
(\textit{column} 2), which represent the effect due to dose, duration response
curves $g_{ij}(t)$ (\textit{column} 3), which represent the effect due to
exposure time, and the expected dose-time interaction function
$h_{ij}(t)$ (\textit{column} 4).

\begin{figure}

\includegraphics{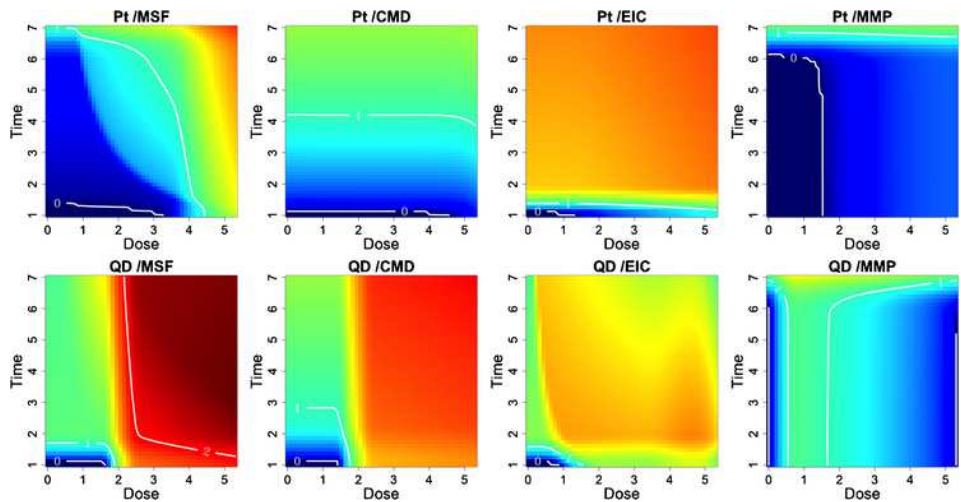}

\caption{Safe exposure regions for the quantum dot (QD) and
platinum (Pt)
nanomaterials. For each particle and outcome we can define dose and time
exposure regions which do not induce cytotoxicity.
Red colored regions indicate greater cytotoxicity to the cells,
whereas blue colored regions indicate reduced risk.
Contour lines quantitate the median estimated response,
relative to the background, where zero response areas can be
interpreted as safe exposure regions.}\vspace*{-3pt}
\label{figsafety}
\end{figure}

\begin{figure}

\includegraphics{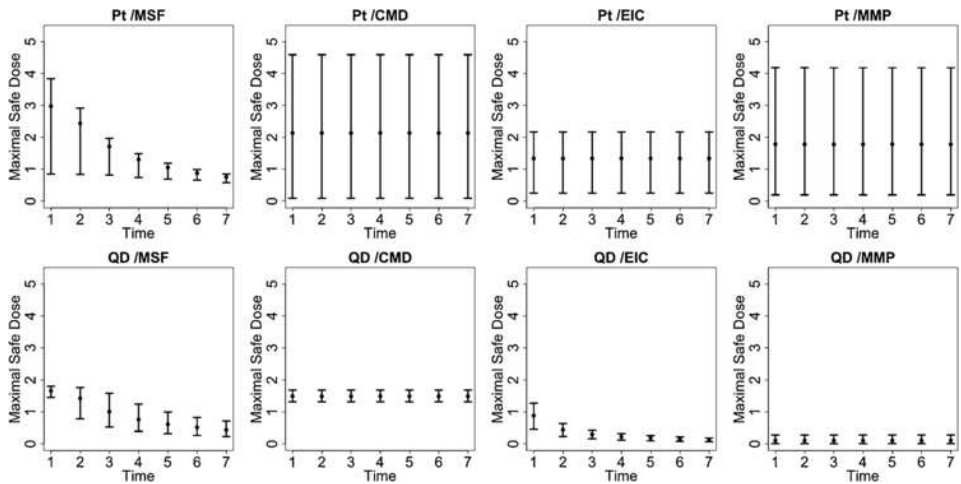}
\caption{Maximal Safe Dose for the quantum dot (QD) and
platinum (Pt) nanomaterials.
Posterior summary estimates of the maximal safe dose, conditional on
exposure time,
including the posterior mean and associated 95\% posterior intervals.
In the case of no interaction, the maximal safe dose is the same across
all times.}
\label{figMSE}
\end{figure}

Figure~\ref{figsafety} provides a plot of the estimated median
response, relative to the background, for different doses and times of
exposure. Blue colors indicate safety regions or areas of reduced risk
to the cells, while red colored regions indicate increased risk of
cytotoxicity. Finally, Figure~\ref{figMSE} provides posterior summary
estimates including mean and $95\%$ posterior intervals for the maximal
safe dose, conditional on the duration of exposure. Note that in the
absence of a dose-time interaction, the maximal safe dose is the same
across all exposure times.

Quantum dot (QD) shows a relatively high toxic response for plasma
membrane damage and mitochondrial superoxide formation. In particular,
we see a more pronounced dose effect for membrane damage and both a
time, dose and significant dose-time interaction ($\hat{\rho}=0.99$)
effect for mitochondrial superoxide formation. This supports what has
previously been demonstrated in conventional assays that QD
nanoparticles stabilized by toluene are capable of inducing tiers 2 and
3 oxidative stress responses induced by the toluene [\citet
{George2011}]. Platinum~(Pt) shows a high dose and time response for
mitochondrial superoxide formation, including a significant dose-time
interaction effect ($\hat{\rho}=0.99$), and a pronounced time effect for
elevated calcium but not for mitochondrial depolarization or membrane
damage, indicating that the particle induced sublethal effects to the
cell without cytotoxicity. The Zinc oxide nanoparticle (ZnO), reported
in the supplemental article, Appendix D, shows a relatively high toxic
response for plasma membrane damage, elevated calcium and mitochondrial
depolarization [\citet{Patel2012}]. In particular, we see a more
pronounced time effect for the elevated calcium and both a time and
dose-response for membrane damage and mitochondrial depolarization.
This again verifies what has previously been demonstrated in
conventional assays, since ZnO nanoparticles are capable of inducing
tiers 2 and 3 oxidative stress responses through Zn$_{2}^{+}$ release
[\citet{GeorgePokhrel2010}]. In contrast, the gold nanoparticle (Al),
also reported in the supplemental article, Appendix D, shows very
little response for all outcomes, indicating that, compared to the
other particles, it has small risk of inducing a sublethal or lethal
cytotoxic response [\citet{Patel2012}].\looseness=-1

\section{Discussion} \label{secDiscussion}
In this article we propose a statistical framework for modeling
dependent dose-response surfaces over multivariate outcomes. The
proposed methodology accounts for dose and duration kinetics jointly
using a flexible model which does not compromise interpretability. We
account for the multivariate nature of the data using the hierarchical
framework and thereby efficiently combine information and borrow
strength across cellular injury patterns. We account for the nonrobust
nature of the data by allowing for particle specific variance inflation,
resulting in a $T$-distributed model for the error structure.

The main challenge associated with the class of models proposed in this
manuscript is finding the right balance between model complexity and
model interpretability. An alternative formulation of the dose-response
surface would seek inference for a general smooth surface
$m_{ij}(d,t)$. However, our simplified approach, based on the
assumptions of additivity and linearity, maintains a very appealing
level of interpretability, allowing for the definition of specific risk
assessment parameters while maintaining an adequate level of flexibility.
A related generalization of the proposed additive framework would
include a more general class of
functional interactions to account for a possible synergistic effect between
dose and duration of exposure.
This would come at the cost of reduced interpretability, but, at the
same time,
could be of clear scientific interest in some contexts.
In this initial modeling effort, we choose to work with a $T$-distributed
error structure
and therefore normalize our response to unconstrain the support so that
it can take on values between $-\infty$ and $\infty$.
An alternative formulation could retain the original scale of the data,
but rather define a generalized multivariate model such that the
outcome distribution can be described using binomial or beta random
quantities. This
extension would require a substantial increase
in computational complexity, with the possible need to consider
numerical or analytical approximations, but it is clearly
worthy of further methodological exploration.

The hierarchical formulation introduced in this article is easily
adapted to
the case where multiple cell lines are used to test for cytotoxicity.
A natural integration strategy would perhaps find motivation
in the meta analytic framework, with information shared between experiments
via the structuring of one extra level in the hierarchy.

Finally, the proposed model can also be expanded by the inclusion of covariates.
This is naturally defined as an extension to stage 3 of the model
introduced in Section~\ref{secModel}.
The addition of covariates is especially important for relating
specific ENM properties to toxicity, and is therefore an important area
for future work.

\section*{Acknowledgments}
Primary support was provided by the U.S. Public Health Service Grant
U19 ES019528 (UCLA Center for Nanobiology and Predictive Toxicology).
This work was also supported by the National Science Foundation and the
Environmental Protection Agency under Cooperative Agreement Number
DBI-0830117. Any opinions, findings, conclusions or recommendations
expressed herein are those of the author(s) and do not necessarily
reflect the views of the National Science Foundation or the
Environmental Protection Agency. This work has not been subjected to an
EPA peer and policy review.

\begin{supplement}[id=suppA]
\stitle{Supplementary Appendices}
\slink[doi]{10.1214/12-AOAS563SUPP} 
\slink[url]{http://lib.stat.cmu.edu/aoas/563/supplement.pdf}
\sdatatype{.pdf}
\sdescription{Full conditional distributions for the model described in Section~\ref
{secModel} are provided in the supplemental article, Appendix A. Spline
coefficients $\bbeta,\bgamma$ and $\bdelta$ are directly sampled
from their conditional posterior distributions via direct simulation
(Gibbs step). To assess estimation of the model presented in Section~\ref{secModel}, we present a simulation study in the supplemental
article, Appendix B. The dose and time kinetics were simulated from
various parametric functions. Both canonical and noncanonical profiles
that are reasonably interpretable under a toxicity framework were
generated. In addition, we assess sensitivity of the model results to
our choice of prior parameters for population level interior knot
parameters $\bmm{\lambda}_{\bolds{\phi}_i}$ and $\bmm{\lambda}_{\bolds{\phi}_i}$.
In the
supplemental article, Appendix~C, we provide an additional sensitivity
analysis assessing model results to our choice of prior model for the
change-point parameters. Alternative prior models assessed include a
truncated normal prior and a parameterization of the bivariate beta
prior that results in a uniform prior on the simplex. The supplemental
article, Appendix D, presents results associated with inference on the
6 remaining particles not presented in Section~\ref{subsecAnalysis}.
Finally, Appendix E discusses model assessment and goodness-of-fit
diagnostics associated with the model described in Section~\ref{secModel}.}
\end{supplement}


%


\printaddresses

\end{document}